\documentclass[showpacs,prd]{revtex4}
\usepackage{graphicx}
\usepackage{dcolumn}
\usepackage{bm}
\begin{document}
\title{Factorization fits and the unitarity triangle in charmless
two-body $B$ decays}
\author{N. \surname{de Groot}}
\affiliation{Rutherford Appleton Laboratory, Chilton Didcot OX11 0QX, UK}
\author{W. N. Cottingham}
\affiliation{H. H. Wills Physics Laboratory, Bristol University,\\
Tyndall Avenue, Bristol BS8 1TL, UK}
\author{I. B. Whittingham}
\affiliation{School of Mathematical and Physical Sciences,\\
James Cook University, Townsville, Australia, 4811}
\date{\today}
\begin{abstract}
We present fits to charmless hadronic $B$ decay data from the BaBar, Belle
and Cleo experiments using two theoretical models: (i) the QCD factorization
model of Beneke \textit{et al.} and (ii) QCD factorization complemented with
the so-called \textit{charming penguins} of Ciuchini \textit{et al.}
When we include the data from
pseudoscalar-vector decays the results favor the incorporation of
the \textit{charming penguin} terms. We also present fit results for
the unitarity triangle parameters and the CP-violating asymmetries.
\end{abstract}

\maketitle

\section{\label{sec:intro} Introduction}

A wealth of experimental data on hadronic charmless $B$ decays has become
available from the BaBar and Belle experiments.
These studies of the numerous $B$ decay channels are designed to test the
Cabbibo-Kobayashi-Maskawa (CKM) explanation of $CP$ violation in the
standard model as represented by the unitarity triangle condition
$V_{ud}V^{*}_{ub}+V_{cd}V^{*}_{cb}+V_{td}V^{*}_{tb}=0$ on the CKM mixing
parameters. Such tests have been facilitated by the recent significant
progress in the theoretical understanding of hadronic decay amplitudes
based upon QCD factorization which allows the amplitudes to be expressed
in terms of numerous soft QCD parameters, such as meson decay constants and
transition form factors, and a set of calculable coefficients.
In this paper we present an analysis of recent data,
based upon QCD factorization. We also investigate
the potential contribution to the decay amplitudes of $b$ quark annihilation
and so-called charming penguins. The data that we attempt to fit includes
decays to
pseudoscalar ($\pi $ and $K$) and vector ($\rho, \omega , K^{*}$ and $\phi$)
mesons. This is an extension of an earlier study \cite{Cott}
that was based upon
simplified formulae derived from the heavy quark limit of QCD factorization.

The starting point for the calculation of all $B$ meson decay amplitudes is
the effective Hamiltonian
\begin{equation}
\label{Heff}
\mathcal{H}_{\text{eff}}(\mu )  = \frac{G_{F}}{\sqrt{2}}
 \sum_{p=u,c}\lambda _{p}
[ C_{1}(\mu )Q^{p}_{1} + C_{2}(\mu )Q^{p}_{2} 
 + \sum_{i=3, \ldots , 10}C_{i}(\mu )Q_{i}]  +
\text{other terms}
\end{equation}
where $\lambda_{p}=V^{*}_{pq}V_{pb}$ is a product of CKM matrix elements,
$q=d,s$ and the local $\Delta B =1$ four-quark operators are
\begin{eqnarray}
\label{4q-ops}
Q^{p}_{1}& =& (\bar{q}_{\alpha}p_{\alpha})_{V-A}
(\bar{p}_{\beta}b_{\beta})_{V-A},  \nonumber  \\*
Q^{p}_{2} & =& (\bar{q}_{\alpha}p_{\beta})_{V-A}
(\bar{p}_{\beta}b_{\alpha})_{V-A}, \nonumber  \\*
Q_{3,5} & =& (\bar{q}_{\beta}b_{\beta})_{V-A}\sum_{q^{\prime}}
(\bar{q}^{\prime}_{\alpha}q^{\prime}_{\alpha})_{V\mp A},  \nonumber  \\*
Q_{4,6} & =& (\bar{q}_{\beta}b_{\alpha})_{V-A}\sum_{q^{\prime}}
(\bar{q}^{\prime}_{\alpha}q^{\prime}_{\beta})_{V\mp A}, \nonumber  \\*
Q_{7,9}& =& (\bar{q}_{\beta}b_{\beta})_{V-A}\sum_{q^{\prime}}
(\bar{q}^{\prime}_{\alpha}q^{\prime}_{\alpha})_{V\pm A}, \nonumber  \\*
Q_{8,10} & =& (\bar{q}_{\beta}b_{\alpha})_{V-A}\sum_{q^{\prime}}
(\bar{q}^{\prime}_{\alpha}q^{\prime}_{\beta})_{V\pm A},
\end{eqnarray}
where $q^{\prime} \in \{u,d,s,c\}$, $\alpha$ and $\beta $ are color indices,
$e_{q^{\prime}}=2/3(-1/3)$ for $u(d)$-type quarks
and we use the notation, for example,
\begin{equation}
Q_{5} = (\bar{q}_{\beta}b_{\beta})_{V-A}\;\sum_{q^{\prime}}
(\bar{q}^{\prime}_{\alpha}q^{\prime}_{\alpha})_{V+A}  
 = \left[\bar{q}_{\alpha}\gamma_{\mu }(1-\gamma_{5})b_{\alpha}\right]
\sum_{q^{\prime}}\left[\bar{q}^{\prime}_{\beta}\gamma^{\mu }(1+\gamma_{5})
q^{\prime}_{\beta}\right].
\end{equation}
$Q_{1,2}$ are the current-current tree operators, $Q_{3,\ldots ,6}$
are QCD penguin operators and $Q_{7,\ldots ,10}$ are electroweak penguin
operators. The 'other terms' indicated in (\ref{Heff})
include the electromagnetic and chromomagnetic dipole transition
operators.  In the Standard Model the contributions from the electroweak
penguin operators and magnetic dipole operators are generally small. The
exception is the electroweak penguin operator
coefficient $C_{9}$ which is larger than the QCD penguin coefficients $C_{3}$
and $C_{5}$.

It has been shown by Beneke \textit{et al.} \cite{BBNS} that, in the heavy
quark limit, the hadronic matrix elements of the four-quark operators
in the amplitudes for non-leptonic $B$ decays into two light mesons $M_{1,2}$
have the form
\begin{equation}
\label{matrix}
\langle M_{1}M_{2}|Q_{i}|B \rangle  =  \langle M_{1}|J_{1 \mu}|B \rangle
\langle M_{2}|J^{\mu }_{2}|0\rangle   
 \left[1+\sum_{n} r_{n}\alpha_{s}^{n} +
O(\Lambda_{\text{QCD}}/m_{b})\right]
\end{equation}
and can be calculated from first principles, including non-factorizable
strong interaction corrections. QCD factorization extends naive
factorization by separating matrix elements into short distance
contributions at scale $O(1/m_{b})$ that are perturbatively
calculable and long distance contributions $O(1/\Lambda_{\text{QCD}})$
that are parameterized.

$B$ meson decay can also be initiated by $b$ quark annihilation with its
partner. Although the annihilation contributions to the decay amplitude
are formally of $O(\Lambda_{\text{QCD}}/m_{b})$ and power suppressed, they
violate QCD factorization because of end point divergences. However these
weak annihilation contributions can be included into the decay amplitudes
by treating the end point divergences as phenomenological parameters.
Analyses based upon QCD factorization with inclusion of weak annihilation
have been undertaken for $B \rightarrow PP$ \cite{BBNS,Du1} and
$B \rightarrow PV$ \cite{Du2}.
Although general agreement with experiment
was found, some branching ratios for $PV$ decays were only marginally
consistent and all predictions were plagued by the large uncertainties
associated with poorly determined parameters within the theory. A recent
global analysis of $PP$ and $PV$ decays \cite{Du3} found that QCD
factorization plus weak
annihilation could fit many decay channels but yielded results too low for the
$B \rightarrow \pi K^{*}$ decays.  More recently, Aleksan \textit{et al.}
\cite{Aleksan} have undertaken a global analysis of $PV$ decays and have
concluded that QCD factorization cannot fit the experimental data when
decay channels involving $K^{*}$ mesons are included, and suggest that this
failure is due to some larger than expected nonperturbative contribution.
Motivated by the concept of so-called charming penguins,
that is non-perturbative $O(\Lambda_{\text{QCD}}/m_{b})$
corrections from enhanced $c$-loop penguins, first introduced by Ciuchini
\textit{et al.} \cite{Ciuchini}, they introduce additional long-distance
contributions to the decay amplitudes and include the two complex
parameters from these additional amplitudes in their global fit. They
obtain a slightly better fit but their best-fit parameters are at the
limits of the allowed domain.
A recent study \cite{Guo}, limited to $B \rightarrow \rho \pi $ decays, has
used QCD factorization to place bounds on the $F^{B \rightarrow \pi}_{1}$
form factor.
A detailed study of QCD factorization applied
to $PP$ and $PV$ decays of $B$ mesons
has just been completed by Beneke and Neubert \cite{BenekePV}. Predicted
branching ratios and $CP$ asymmetries for a large number of $PP$ and $PV$
channels are given for default values of input parameters
and detailed estimates of the
theoretical uncertainties in these predictions determined for various
scenarios of input parameters. Beneke and Neubert find that there is a
scenario for which there is general global agreement between the
results of QCD factorization and measurement except for
$\overline{B}^{0} \rightarrow \pi^{0}\overline{K}^{0}$ and the group of
$B \rightarrow \pi K^{*}$ decays.

In this paper we undertake a global analysis of 18 $PP$, $PV$ and
$VV$ channels
using two theoretical models, (i) QCD factorization with inclusion of
weak annihilation and (ii) this model supplemented with charming penguins.
Our study is similar in spirit to that of
Aleksan \textit{et al.} \cite{Aleksan} for $PV$ decays
but we extend the global
fit to include $PP$ and some $VV$ decays.
This paper is organized as follows. In
Sec. II we review the structure of the decay amplitude within QCD
factorization and discuss the various parameters that occur in this
amplitude. Inclusion of weak annihilation and charming penguins is
discussed in Sec. III and Sec. IV respectively. The method and results of
our best fit for our two models to current experimental data is presented in
Sec V, and Sec. VI contains our discussion and conclusions.

\section{DECAY AMPLITUDE IN QCD FACTORIZATION}

In QCD factorization, the amplitude for $B$ decay into two
light hadrons (mesons) $M_{1,2}$ has the form, neglecting weak
annihilation processes,
\begin{eqnarray}
\label{decay-amp}
\langle M_{1}M_{2}|\mathcal{H}_{\text{eff}}|B \rangle & =
& \frac{G_{F}}{\sqrt{2}}
\{\sum_{i=1,2} \lambda_{u}a^{u}_{i} 
[T_{i}(M_{1},M_{2})+T_{i}(M_{2},M_{1})]  \nonumber  \\*
& & +\sum_{p=u,c}\sum_{i=3,\ldots ,6,9} \lambda_{p}a^{p}_{i} 
[T_{i}(M_{1},M_{2}) +T_{i}(M_{2},M_{1})] \}
\end{eqnarray}
where
\begin{eqnarray}
\label{t-amps}
T_{1}(M_{1},M_{2})&=&\langle M_{1}|\bar{u}\gamma^{\mu}
(1-\gamma_{5})b|B\rangle 
 \langle M_{2}|\bar{q}\gamma_{\mu }(1-\gamma_{5})u|0 \rangle,
\nonumber  \\
T_{2}(M_{1},M_{2})&=&\langle M_{1}|\bar{q}\gamma^{\mu}
(1-\gamma_{5})b|B\rangle 
 \langle M_{2}|\bar{u}\gamma_{\mu }(1-\gamma_{5})
u|0 \rangle,  \nonumber  \\
T_{3}(M_{1},M_{2})&=&\langle M_{1}|\bar{q}\gamma^{\mu}
(1-\gamma_{5})b|B\rangle 
\langle M_{2}|\bar{q}^{\prime}\gamma_{\mu }
(1-\gamma_{5})q^{\prime}|0 \rangle,  \nonumber  \\
T_{4}(M_{1},M_{2})&=&\langle M_{1}|\bar{q}^{\prime}\gamma^{\mu}
(1-\gamma_{5})b|B\rangle 
 \langle M_{2}|\bar{q}\gamma_{\mu }
(1-\gamma_{5})q^{\prime}|0 \rangle,  \nonumber  \\
T_{5}(M_{1},M_{2})&=&\langle M_{1}|\bar{q}\gamma^{\mu}
(1-\gamma_{5})b|B\rangle 
 \langle M_{2}|\bar{q}^{\prime}\gamma_{\mu }
(1+\gamma_{5})q^{\prime}|0 \rangle,  \nonumber  \\
T_{6}(M_{1},M_{2})&=& -2\langle M_{1}|\bar{q}^{\prime}
(1-\gamma_{5})b|B\rangle 
 \langle M_{2}|\bar{q}(1+\gamma_{5})q^{\prime}
|0\rangle,  \nonumber  \\
T_{9}(M_{1},M_{2})&=&\langle M_{1}|\bar{q}\gamma^{\mu}
(1-\gamma_{5})b|B\rangle   
\langle M_{2}|e_{q^{\prime}}\bar{q}^{\prime}
\gamma_{\mu }(1-\gamma_{5})q^{\prime}|0 \rangle .
\end{eqnarray}
The two-quark matrix elements are the well determined electroweak decay
constants $f_{\pi},f_{K},f_{\rho},$ e.t.c. and the $B$ transition form
factors $F_{\pi},F_{K},A_{\rho},$ e.t.c. In principle transition
form factors are independently measurable in $B$ semi-leptonic decays but
to date they are only loosely constrained by measurements and model
estimations.

The coefficients $a_{i}$ have to be calculated
from the Wilson coefficients \cite{BBNS}.
We have calculated the Wilson coefficients at several scales $\mu $
of
$O(m_{b})$ using the next-to-leading-order (NLO) renormalization group
equations
\begin{equation}
\label{rges}
C_{i}(\mu ) = U_{ij}(\mu ,M_{W})C_{j}(M_{W}).
\end{equation}
We follow the Beneke \textit{et al.} \cite{BBNS} prescription of (i)
dropping terms of $O(\alpha_{s}^{2})$, $O(\alpha^{2})$ and
$O(\alpha_{s} \alpha)$ in (\ref{rges}), (ii) neglecting the effect of the
electromagnetic penguins $C_{7, \ldots ,10}(M_{W})$ on the evolution
of the QCD penguin coefficients $C_{1, \ldots ,6}$ and, (iii) in
$C_{j}(M_{W})$, splitting the $O(\alpha )$ electroweak penguin terms
into those enhanced by large $m_{t}$ or $1/\sin ^{2}\theta_{W}$,
which are treated as leading order (LO), and treating the remainder,
together with the $O(\alpha_{s})$ terms, as NLO. Our calculated values
are shown in tables \ref{CLO} and \ref{CNLO} 
and are very similar to those of Beneke
\textit{et al.} \cite{BBNS} and Du \textit{et al.} \cite{Du1,Du2}.

\begin{table}
\caption{\label{CLO} Leading order Wilson coefficients $C_{i}$ in the NDR
scheme calculated at the scales $\mu $ and $\mu_{h}=\sqrt{\Lambda_{h} \mu }$,
where $\Lambda_{h}=0.5$ GeV, for $\mu =m_{b}$ and $\mu =m_{b}/2$.
The input parameters are
$\Lambda ^{\overline{MS}(5)}_{\text{QCD}}=0.225$ GeV, $m_{t}(m_{t})=167.0$
GeV, $m_{b}(m_{b})=4.2$ GeV, $M_{W}=80.42$ GeV, $\alpha =1/129$, and
$\sin ^{2} \theta_{W}= 0.23$.}
\begin{ruledtabular}
\begin{tabular}{ldddd}
Scale (GeV) & 4.20 & 2.10  & 1.45 & 1.02 \\
\hline
$C_{1}$ & 1.1174 & 1.1848 & 1.2392 & 1.3120  \\
$C_{2}$ & -0.2678  & -0.3873 & -0.4755 & -0.5862  \\
$C_{3}$ & 0.0121  & 0.0185  & 0.0235  & 0.0299  \\
$C_{4}$ & -0.0274  & -0.0383 & -0.0459 & -0.0551  \\
$C_{5}$ & 0.0080  & 0.0105  & 0.0120  & 0.0136  \\
$C_{6}$ & -0.0341  & -0.0526  & -0.0677 & -0.0883  \\
$C_{7}/\alpha$ & -0.0140  & -0.0282  & -0.0314 & -0.0303  \\
$C_{8}/\alpha$ & 0.0288  & 0.0432  & 0.0555 & 0.0734  \\
$C_{9}/\alpha$ & -1.2913  & -1.3785  & -1.4408  & -1.5190 \\
$C_{10}/\alpha$ & 0.2888  & 0.4213  & 0.5194  & 0.6424 \\
\end{tabular}
\end{ruledtabular}
\end{table}

\begin{table}
\caption{\label{CNLO} Next to leading order
Wilson coefficients $C_{i}$ in the NDR
scheme calculated at the scales $\mu $ and $\mu_{h}=\sqrt{\Lambda_{h} \mu }$,
where $\Lambda_{h}=0.5$ GeV, for $\mu =m_{b}$ and $\mu =m_{b}/2$.
The input parameters are
$\Lambda ^{\overline{MS}(5)}_{\text{QCD}}=0.225$ GeV, $m_{t}(m_{t})=167.0$
GeV, $m_{b}(m_{b})=4.2$ GeV, $M_{W}=80.42$ GeV, $\alpha =1/129$, and
$\sin ^{2} \theta_{W}= 0.23$.}
\begin{ruledtabular}
\begin{tabular}{ldddd}
Scale (GeV) & 4.20 & 2.10  & 1.45 & 1.02 \\
\hline
$C_{1}$ & 1.0813 & 1.1374 & 1.1820 & 1.2405  \\
$C_{2}$ & -0.1903  & -0.2948 & -0.3700 & -0.4619  \\
$C_{3}$ & 0.0137  & 0.0212  & 0.0274  & 0.0358  \\
$C_{4}$ & -0.0357  & -0.0506 & -0.0618 & -0.0762  \\
$C_{5}$ & 0.0087  & 0.0102  & 0.0105  & 0.0096  \\
$C_{6}$ & -0.0419  & -0.0653  & -0.0854 & -0.1146  \\
$C_{7}/\alpha$ & -0.0026  & -0.0139  & -0.0147 & -0.0100  \\
$C_{8}/\alpha$ & 0.0618  & 0.0986  & 0.1317 & 0.1813  \\
$C_{9}/\alpha$ & -1.2423  & -1.3181  & -1.3526  & -1.4149 \\
$C_{10}/\alpha$ & 0.2283  & 0.3388  & 0.4015  & 0.4991 \\
\end{tabular}
\end{ruledtabular}
\end{table}

To lowest order in the strong coupling constant $\alpha_{s}$, the
$a_{i}$ coefficients are the
same as in naive factorization, that is
\begin{equation}
\label{aLO}
a^{\text{LO}}_{i} = C_{i} + C_{i^{\prime}}/N_{c}
\end{equation}
where $i^{\prime}=i-(-1)^{i}$ and $N_{c}=3$ is the number of quark colors.
The higher order corrections include
lowest order gluon exchange between the quarks in the basic tree amplitudes
which are calculated and folded into the light cone distribution functions
$\Phi_{M}(x,\mu )$ of the participating mesons
(see, for example, \cite{Ball}). This results in the $a_{i}$ coefficients
having the form
\begin{equation}
\label{acofs}
a_{i}(M_{1}M_{2})=a_{i,I}(M_{2})+a_{i,II}(M_{1}M_{2})
\end{equation}
where $M_{1}$ is the recoil meson containing the spectator (anti) quark
and $M_{2}$ is the emitted meson. The complex quantities $a_{i,I}$ describe
the formation of $M_{2}$, including nonfactorizable corrections from
hard gluon exchange or light quark loops in penguins. They do not involve
the hard gluon exchanges with the spectator quark, these are described by the
(possibly) complex quantities $a_{i,II}$.

In the correction terms the leading-twist light cone distribution
functions for both
pseudoscalar and vector mesons are expanded in the first few terms of a
Gegenbauer expansion
\begin{equation}
\label{Gegen}
\Phi_{M}(x,\mu ) = 6x (1-x) [1+\sum_{n} \alpha^{M}_{n} C^{3/2}_{n}(2x-1)].
\end{equation}
The asymptotic limit $\Phi_{M}(x,\mu )=6x(1-x)$ is valid for the mass
scale $\mu \rightarrow \infty $. The parameters $\alpha^{M}_{n}$ are
anticipated to be small but they are not well established. To economize in
the number of fitting parameters in the initial fits to data
presented here they are taken to be zero. With this simplification all
light mesons included in our analysis have the same spatial wavefunction
and all coefficients $a_{i,I}$ except $a_{6,I}$ are the same for all decays.
Formulae for
the evaluation of the $a_{i,I}$ can be found in \cite{BBNS,Du2,BenekePV}.
Although Beneke and Neubert \cite{BenekePV} obtain a different expression
for $a_{6,I}(M_{2}=V)$ to that of Du \textit{et al.} \cite{Du2}, this
is not important here as $a_{6}(PV)$ does not  occur in the decay amplitudes
for $M_{2}=V$ since $\langle V|(\bar{q}q)_{S\pm P}|0 \rangle =0$.
The $O(\alpha_{s})$ corrections to $a^{\text{LO}}_{i}$ include contributions
from one-loop vertex corrections $V_{M}(\mu ,\alpha^{M}_{n})$ to all
$a_{i,I}$ and from penguin
corrections $P_{M}(\mu , \alpha^{M}_{n},s_{p})$, involving
the quark mass ratio $s_{p}=(m_{p}/m_{b})^{2}$, to $a_{4,I}$ and $a_{6,I}$.
We neglect the small electroweak penguin corrections to $P_{M}$. 
Typical parton off-shellness in the loop
diagrams contributing to $V_{M}$ and $P_{M}$ is
$O(m_{b})$, suggesting that the scale $\mu \propto m_{b}$ should be used
in evaluating $a_{i,I}$. The results
of our calculations of $a_{i,I}$ coefficients without light cone
corrections are given in table \ref{aiI0}. Results including the light cone
corrections $\alpha^{M}_{1,2}$ taken from \cite{LuYang} are given in
table \ref{aiI1}. The coefficients most affected are $a_{2,I}$
and $a_{4,I}$.

\begin{table*}
\caption{\label{aiI0} Factorization $a_{i,I}(M_{2})$
coefficients evaluated at the scale $\mu $ using the expressions of
\cite{BBNS,BenekePV}. The light cone corrections $\alpha^{M_{2}}_{n}$
are set equal to
zero, making all coefficients except $a_{6,I}$ universal.}
\begin{ruledtabular}
\begin{tabular}{cdd}
$\mu $(GeV) & 4.20  & 2.10 \\
\hline
$a^{u}_{1,I}$ & 1.0572 + 0.0200i & 1.0791+0.0369i \\
$a^{u}_{2,I}$ & 0.0060-0.0836i   & -0.0377-0.1130i  \\
$a_{3,I}$     & 0.0058+0.0021i   &  0.0083+0.0037i  \\
$a^{u}_{4,I}$ & -0.0312-0.0161i  & -0.0338-0.0205i  \\
$a^{c}_{4,I}$ & -0.0369-0.0068i  & -0.0415-0.0079i  \\
$a_{5,I}$     & -0.0070-0.0026i  & -0.0106-0.0050i  \\
$a^{u}_{6,I}(P)$ & -0.0433-0.0152i  & -0.0586-0.0188i  \\
$a^{c}_{6,I}(P)$ & -0.0465-0.0056i  & -0.0630-0.0056i  \\
$a^{u}_{6,I}(V)$ & -0.0075-0.0007i  & -0.0094-0.0013i  \\
$a^{c}_{6,I}(V)$ & 0.0009-0.0115i  & 0.0019+0.0152i  \\
$a_{9,I}$     & -0.0094-0.0002i  & -0.0097-0.0003i  \\
\end{tabular}
\end{ruledtabular}
\end{table*}
\begingroup
\squeezetable
\begin{table*}
\caption{\label{aiI1} Factorization $a_{i,I}(M_{2})$
coefficients evaluated at the scale $\mu =2.10$ GeV using the
expressions of \cite{BBNS,BenekePV} and L\"{u} and Yang
\cite{LuYang} values for
the light cone corrections $\alpha^{M_{2}}_{1,2}$.}
\begin{ruledtabular}
\begin{tabular}{cdddd}
    & \multicolumn{1}{c}{$\pi $}  & \multicolumn{1}{c}{$K$}  &
    \multicolumn{1}{c}{$\rho,\omega $} & \multicolumn{1}{c}{$K^{*}$}  \\

\hline
$a^{u}_{1,I}$ & 1.0809 + 0.0369i & 1.0762+0.0432i & 1.0798+0.0369i &
1.0753 +0.0439i  \\
$a^{u}_{2,I}$ & -0.0432-0.1130i  & -0.0290-0.1322i & -0.0400-0.1130i
&  -0.0261-0.1344i  \\
$a_{3,I}$ & 0.0085+0.0037i  &  0.0080+0.0043i  & 0.0084+0.0037i
& 0.0079 +0.0043i \\
$a^{u}_{4,I}$ & -0.0306-0.0206i & -0.0313-0.0210i & -0.0325-0.0206i
&  -0.0321-0.0210i  \\
$a^{c}_{4,I}$ & -0.0343-0.0072i  & -0.0364-0.0061i & -0.0386-0.0076i
&  -0.0383-0.0061i  \\
$a_{5,I}$ & -0.0108-0.0050i  & -0.0102-0.0059i & -0.0107-0.0050i
&  -0.0101-0.0060i  \\
$a^{u}_{6,I}$ & -0.0586-0.0188i  & -0.0586-0.0188i & -0.0097-0.0012i
&  -0.0102-0.0013i  \\
$a^{c}_{6,I}$  & -0.0630-0.0058i  &  -0.0630-0.0058i & 0.0022+0.0147i
&  0.0032+0.0152i  \\
$a_{9,I}$ & -0.0097-0.0003i  & -0.0097-0.0004i & -0.0097-0.0003i
&  -0.0097-0.0004i  \\
\end{tabular}
\end{ruledtabular}
\end{table*}
\endgroup

The coefficients $a_{i,II}$ are not universal even when light cone
corrections are neglected. They contain not only the low energy parameters
(decay constants and form factors) from the lowest order
calculations, in common with naive factorization, but low energy contributions
to the folding integrals involving another non-perturbative complex
parameter $X_{H}$ which is only loosely constrained by model estimations.
To discuss the form of the $a_{i,II}$, we first note that, from (\ref{Heff}),
the contributions of the $a_{i}$ coefficients
to the decay amplitude for $B \rightarrow M_{1}M_{2}$ are of the form
\begin{equation}
\label{ai-contrib}
m_{B}^{2}\frac{G_{F}}{\sqrt{2}}\lambda_{p}[g^{i}_{1,2}f_{M_{1}}F_{M_{2}}
+g^{i}_{2,1}f_{M_{2}}F_{M_{1}}]a_{i},
\end{equation}
where $g^{i}_{1,2}$ and $g^{i}_{2,1}$ are products of Clebsch-Gordan
coefficients tablulated in, for example, \cite{Cott,AliGreub}
\footnote{It should be noted that these tables conform to the sign
conventions of Ali and Greub \cite{AliGreub} and differ from the isospin
convention of Beneke \textit{et al.} \cite{BBNS}}. Using the formulae
of \cite{BBNS}, we can write, with $N_{c}=3$,
\begin{equation}
\label{aiII}
f_{M_{2}}F_{M_{1}}a_{i,II} = \frac{4 \pi }{9}\epsilon_{i}C_{i^{\prime}}
\alpha_{s} \beta_{i}
\end{equation}
where $\epsilon_{i}=+1 (i=1, \ldots ,4,9), \epsilon_{5}=-1,
\epsilon_{6}=0$,  and
\begin{eqnarray}
\label{betai}
\beta_{i} & = & \frac{f_{B}f_{M_{2}}f_{M_{1}}}{m_{B}\lambda_{B}}
\left[3(1+\epsilon_{i}\alpha^{M_{2}}_{1}+\alpha^{M_{2}}_{2})
(1+\alpha^{M_{1}}_{1}+\alpha^{M_{1}}_{2}) \right. \nonumber  \\
&& + \left. r^{M_{1}}_{\chi }(1-\epsilon_{i}\alpha^{M_{2}}_{1}+
\alpha^{M_{2}}_{2})X^{M_{1}}_{H})\right] .
\end{eqnarray}
The chiral factors $r^{M_{1}}_{\chi}$ are zero for $M_{1}$ a vector
meson and are
\begin{equation}
\label{chiral}
r^{\pi}_{\chi}=\frac{2 m_{\pi}^{2}}{m_{b}(m_{u}+m_{d})}, \quad
r^{K}_{\chi}=\frac{2 m_{K}^{2}}{m_{b}(m_{u}+m_{s})}
\end{equation}
for the pseudoscalar mesons.
It should be noted that these $a_{i,II}$ contributions to the decay
amplitudes are independent of the $B$ transition form factors. However,
they do involve the poorly determined parameter $f_{B}/\lambda_{B}$ where
$f_{B}$ is the $B$ leptonic decay constant and $\lambda_{B} \approx 0.6$
GeV is related to the $B$ light cone distribution function.
The parameter $X^{M_{1}}_{H}$ is the contribution of a logarithmic
end-point divergence in the integration over the $M_{1}$ light cone
distribution function
\begin{equation}
\label{XH}
X^{M_{1}}_{H} = \int ^{1}_{0}\frac{dx}{1-x}.
\end{equation}
These functions take no account of the internal quark transverse momenta
which, if included, would make the integrals finite but not calculable
within perturbative QCD. $X^{M_{1}}_{H}$ is parameterized as
\begin{equation}
\label{XH-par}
X^{M_{1}}_{H}=\ln \left( \frac{m_{B}}{\Lambda_{\text{QCD}}}\right)
+ \rho_{H}e^{i\phi_{H}}
\end{equation}
where $\rho_{H}$ is not expected to be larger than 3. We take
$\ln (m_{B}/\Lambda_{\text{QCD}})=3.03$. The energies
involved in the calculation of $a_{i,II}$ imply that the appropriate
scale is not that of the scale $\mu $ used in calculating the $a_{i,I}$
but $\mu _{h} = \sqrt{\Lambda_{h} \mu }$ where \cite{BBNS}
$\Lambda_{h}=0.5$ GeV. We use this with $\mu =m_{b}/2$ in our fitting so that
$\alpha_{s}f_{B}/(m_{B}\lambda_{B})=0.0209$.  We note that substantial
light cone corrections $\alpha^{M}_{1,2}$ can significantly enhance
the $a_{i,II}$ coefficients.

\section{Annihilation contributions}

Because of the heavy $b$ quark mass it is expected that perturbative QCD
calculations will give a reliable estimate of the annihilation contribution
to the decay amplitude. In these calculations the basic perturbative quark
amplitudes are again folded into the participating meson light cone
distribution functions. Apart from the low energy regions of the folding
integrals the only low energy parameters that appear in the lowest order
calculations are the participating meson electroweak decay constants
$f_{B}, f_{\pi}, f_{\rho},$ e.t.c. Again the low energy contributions
to the integral introduce another non-perturbative complex parameter
$X_{A}$, which is only loosely constrained by the model estimations. Detailed
formulae are to be found in \cite{BBNS,Du2}. In this paper we follow the
more extensive calculations of Du \textit{et al.} \cite{Du2} but express
the annihilation contribution to the decay amplitude in the form
\begin{eqnarray}
\label{annih}
\langle M_{1}M_{2}|\mathcal{H}^{\text{ann}}_{\text{eff}}|B \rangle & = &
\frac{G_{F}}{\sqrt{2}}B_{M_{1}M_{2}} \{\lambda_{u}
[d_{1}C_{1}+d_{2}C_{2}]A^{i}_{1} 
-\lambda_{t}\{d_{3}[C_{3}A^{i}_{1}+C_{5}A^{i}_{3}]  \nonumber  \\*
&&+[d_{4}C_{4}+d_{6}C_{6}]A^{i}_{1} 
+d_{5}[C_{5}+N_{c}C_{6}]A^{f}_{3}\}
\end{eqnarray}
where
\begin{equation}
\label{Bcoeff}
B_{M_{1}M_{2}}= \frac{C_{F}}{N_{c}^{2}}f_{B}f_{M_{1}}f_{M_{2}}
\end{equation}
and $C_{F}=(N_{c}^{2}-1)/2N_{c}$. The quantities
$A^{i,f}_{1,3}(M_{1},M_{2})$, where the superscript $i(f)$ denotes gluon
emission from initial (final) state quarks, result from
folding the quark amplitudes into the meson distribution functions. If the
asymptotic form of (\ref{Gegen}) is used then \cite{Du2}
\begin{eqnarray}
\label{A-cofs}
A^{i}_{1}(P_{1},P_{2})&=&\pi \alpha_{s}
\left[18\left(X_{A}-4+\frac{\pi^{2}}{3}\right)+2r^{P_{1}}_{\chi}
r^{P_{2}}_{\chi}X^{2}_{A}\right],  \nonumber  \\
A^{i}_{3}(P_{1},P_{2})&=& 6 \pi \alpha_{s}(r^{P_{1}}_{\chi}-
r^{P_{2}}_{\chi})(X^{2}_{A}-2X_{A}+\frac{\pi^{2}}{3}), \nonumber  \\
A^{f}_{3}(P_{1},P_{2}) &=& 6 \pi \alpha_{s}
(r^{P_{1}}_{\chi}+r^{P_{2}}_{\chi})(2X_{A}^{2}-X_{A}), \nonumber  \\
A^{i}_{1}(P,V)&=& \pi \alpha_{s}
\left[18\left(X_{A}-4+\frac{\pi^{2}}{3}\right)\right],  \nonumber  \\
A^{i}_{3}(P,V)& =& \pi \alpha_{s}r^{P}_{\chi}[2\pi^{2}+6(X_{A}^{2}-2X_{A})],
\nonumber  \\
A^{f}_{3}(P,V)&=& 6 \pi \alpha_{s} r^{P}_{\chi}(2X_{A}^{2}-X_{A}),
\nonumber  \\
A^{i}_{1}(V_{1},V_{2}) & = & A^{i}_{1}(P,V),  \nonumber  \\
A^{i}_{3}(V_{1},V_{2})& = & A^{f}_{3}(V_{1},V_{2})=0.
\end{eqnarray}
The coefficients $d_{i}(M_{1},M_{2})$ are Clebsch-Gordan type factors and
are given in table \ref{dcofs} for the particle sign conventions of
\cite{Cott,Du1,Du2,AliGreub}. Note that, in using (\ref{annih}) with
table \ref{dcofs}, there is no need to distinguish between $PV$ and
$VP$ decays.

\begin{table}
\caption{\label{dcofs} Annihilation coefficients $d_{i}(M_{1},M_{2})$ for
$B \rightarrow M_{1}M_{2}$. The VV channels refer to zero helicity states
only.}
\begin{ruledtabular}
\begin{tabular}{ccccccc}
$M_{1}M_{2}$ & $d_{1}$ & $d_{2}$ & $d_{3}$ & $d_{4}$ & $d_{5}$ & $d_{6}$  \\
\hline
$\pi^{+}\pi^{-}$ & $-1$ & 0 & $-1$ & $-2$ & $-1$ & $-2$  \\
$\pi^{0}\pi^{0}$ & $-1$ & 0 & $-1$ & $-2$ & $-1$ & $-2$  \\
$\pi^{0}\pi^{-}$ & 0 & 0 & 0 & 0 & 0 & 0 \\
$\rho^{+}\pi^{-}$ & 1 & 0 & 1 & 2 & $-1$ & $-2$  \\
$\rho^{-}\pi^{+}$ & 1 & 0 & 1 & 2 & 1 & $-2$  \\
$\rho^{0}\pi^{0}$ & 1 & 0 & 1 & 2 & 0 & $-2$ \\
$\rho^{0}\pi^{-}$ & 0 & 0 & 0 & 0 & $-\sqrt{2}$ & 0  \\
$\rho^{-}\pi^{0}$ & 0 & 0 & 0 & 0 & $\sqrt{2}$ & 0  \\
$\omega \pi^{-}$ & 0 & $\sqrt{2}$ & $\sqrt{2}$ & 0 & 0 & 0 \\
$\rho^{0}\rho^{0}$ & 1 & 0 & 1 & 2 & $-1$ & 2 \\
$\rho^{0}\rho^{-}$ & 0 & 0 & 0 & 0 & 0 & 0  \\
$\omega K^{-}$ & 0 & $1/\sqrt{2}$ & $1/\sqrt{2}$ & 0 & $-1/\sqrt{2}$ & 0 \\
$\omega \overline{K}^{0}$ & 0 & 0 & $1/\sqrt{2}$ & 0 & $-1/\sqrt{2} $ & 0  \\
$\pi^{0}K^{-}$ & 0 & $-1/\sqrt{2}$ & $-1/\sqrt{2}$ & 0 & $-1/\sqrt{2}$ & 0  \\
$\pi^{-}\overline{K}^{0}$ & 0 & $-1$ & $-1$ & 0 & $-1$ & 0  \\
$\pi^{-}\overline{K}^{*0}$ & 0 & 1 & 1 & 0 & 1 & 0  \\
$\pi^{0}\overline{K}^{0}$ & 0 & 0 & $1/\sqrt{2}$ & 0 & $1/\sqrt{2}$ & 0  \\
$\rho^{0}K^{-}$ & 0 & $1/\sqrt{2}$ & $1/\sqrt{2}$ & 0 & $-1/\sqrt{2}$ & 0 \\
$\rho^{0}K^{*-}$ & 0 & $1/\sqrt{2}$ & $1/\sqrt{2}$ & 0 & $-1/\sqrt{2}$ & 0  \\
$\pi^{+}K^{-}$ & 0 & 0 & $-1$ & 0 & $-1$ & 0  \\
$\pi^{+}K^{*-}$ & 0 & 0 & 1 & 0 & 1 & 0  \\
$\rho^{+}K^{-}$ & 0 & 0 & 1 & 0 & $-1$ & 0 \\
$\phi K^{-}$ & 0 & 1 & 1 & 0 & 1 & 0  \\
$\phi K^{*-}$ & 0 & 1 & 1 & 0 & $-1$ & 0 \\
$\phi \overline{K}^{0}$ & 0 & 0 & 1 & 0 & 1 & 0  \\
$\phi \overline{K}^{*0} $ & 0 & 0 & 1 & 0 & $-1$ & 0  \\
\end{tabular}
\end{ruledtabular}
\end{table}

\section{Charming penguins}

In our attempts to fit the data set within the scheme outlined above we
found that the $PP$ branching ratios could be accomodated with acceptable
values of the CKM parameters and transition form factors. However, data
on the $\pi K^{*}$ channels is consistently too large to be accounted for.
The problem is that the $\pi K$ branching ratios are only marginally
larger than the $\pi K^{*}$ ratios. In the QCD factorization scheme
described above the penguin operators $Q_{4}$ and $Q_{6}$ contribute
coherently and almost equally and dominate the $\pi K$ decay amplitudes
whereas $Q_{6}$ is missing from the amplitude for $\pi K^{*}$ decay. This
results in the predicted ratio of the $\pi K$ and $\pi K^{*}$ branching
fractions being too small. Perhaps, staying within the QCD factorization
scheme, this failing can be removed by taking radically different light
cone distribution functions for the $K$ and $K^{*}$ mesons. We investigate
the possibility of significant additional contributions from so-called
\textit{charming penguins} \cite{Ciuchini}.

The largest term in the effective Hamiltonian that produces a strange
quark comes from
\begin{equation}
\label{strange}
{\cal H}  =  \frac{G_{F}}{\sqrt{2}}V^{*}_{cs}V_{cb}[C_{1}(\bar{c}_{\beta}
b_{\beta})_{V-A}(\bar{s}_{\alpha}c_{\alpha})_{V-A}  
 +C_{2}(\bar{s}_{\alpha}b_{\alpha})_{V-A}(\bar{c}_{\beta}c_{\beta})_{V-A}].
\end{equation}
The charming penguins originate from these terms when the $c$ and $\bar{c}$
quarks annihilate. In the general description of two body $B$ decays
as given by Buras and Silvestrini \cite{Buras}, charming penguins have
the topologies $CP_{1}$ and $DP_{2}$ of connected and disconnected
penguins respectively. Ciuchini
\textit{et al.} \cite{Ciuchini}
consider the contribution of these terms to $\pi \pi$ and
$\pi K$ decays in the $SU(3)$ limit. In their notation, and including the
small contribution from the $u$ quark loop, this results in a 
contribution to the decay amplitudes which they express as
\begin{eqnarray}
\label{A-charming}
A_{\pi K}&=& -m_{B}^{2}[V_{tb}V_{ts}^{*}\;\overline{P_{1}}
+ V_{ub}V_{us}^{*}\;\overline{P^{GIM}_{1}}]g, \nonumber  \\
A_{\pi \pi}&=& -m_{B}^{2}[V_{tb}V_{td}^{*}\;\overline{P_{1}}
+ V_{ub}V_{ud}^{*}\;\overline{P^{GIM}_{1}}]g,
\end{eqnarray}
where$\overline{P_{1}}$ and $\overline{P^{GIM}_{1}}$
are two complex numbers that are
independent of the particular channel, $\pi \pi$ or $\pi K$. The only channel
dependence is through the Clebsch-Gordan factor $g$ which is the same as the
Clebsch-Gordan factor in the $a_{4}$ contribution from QCD factorization.
Ciuchini \textit{et al.} suggest that all chirally suppressed terms should
be dropped and replaced with this term.

We take the charming penguin contribution to be from
the penguin topology but to be in addition to the QCD factorization of the
penguin. However, to retain the notation of \cite{Ciuchini}, we express
$\overline{P_{1}}$ and $\overline{P^{GIM}_{1}}$ as
\begin{eqnarray}
\label{p-bar}
\overline{P_{1}} & = &\frac{G_{F}}{\sqrt{2}}(f_{\pi}F_{\pi})
|\overline{D}|e^{i\phi},   \nonumber  \\*
\overline{P^{GIM}_{1}} & = &\frac{G_{F}}{\sqrt{2}}(f_{\pi}F_{\pi})
|\overline{D^{GIM}}|e^{i\phi_{GIM}}.
\end{eqnarray}
With the factor $(f_{\pi}F_{\pi})$, taken here to be $0.042$ GeV, the
dimensionless parameters $|\overline{D}|$ and $|\overline{D^{GIM}}|$
must be less
than unity as the charming penguins are expected to be small
$O(\Lambda_{\text{QCD}}/m_{B})$ corrections.

This simple model must be extended to include vector mesons. To this end
we note that (i) in $PV$ decays the vector meson must have zero helicity and
that, for example, the $s$ and $\bar{q}$ quarks forming the decay
meson $M$ can be expected to
have zero spin projection in their direction of motion irrespective of
whether they form a pseudoscalar or vector meson and (ii) that, when folded
into the same light cone distribution functions, the amplitudes would
be the same. This most simple model extends the $SU(3)$ symmetry to $SU(6)$.
With this \textit{albeit} simple extension, the charming penguin
contribution to a particular amplitude is obtained from the factorization
contribution by reference to the $a_{4,I}$ term. For example, the decay
amplitude for $\overline{B}^{0} \rightarrow \pi^{+}\rho^{-}$ contains a term
\begin{equation}
\label{pi-rho}
-m_{B}^{2}\frac{G_{F}}{\sqrt{2}}f_{\rho}F_{\pi}[V_{ub}V_{ud}^{*}(a^{c}_{4,I}
-a^{u}_{4,I}) + V_{tb}V_{td}^{*}\;a^{c}_{4,I}] .
\end{equation}
The charming penguin contribution is obtained by replacing this with
\begin{equation}
\label{cp-contr}
-m_{B}^{2}[V_{ub}V_{ud}^{*}\;\overline{P^{GIM}_{1}}
+ V_{tb}V_{td}^{*}\;\overline{P_{1}}].
\end{equation}

In the charming penguin amplitude, the $s$ quark is produced
from a left-handed field and can be expected to have predominantly negative
helicity. The $q\bar{q}$ pair emanates from either right-handed or left-handed
fields and, with zero helicity for the produced $(s\bar{q})$ meson, we
expect that the left-handed contribution will dominate to form an $a_{4}$
type term. The right-handed term will be suppressed by a factor
$\Lambda_{\text{QCD}}/m_{B}$. We appreciate that the expectation of
considerable suppression is false for the corresponding factorization
term $a_{6}Q_{6}$ for which the suppression factor $r^{K}_{\chi}$ is of
order unity.  $a_{6}Q_{6}$ is a product of matrix elements of local
operators containing $\langle M|(\bar{s}q)|0 \rangle $ which is zero for
$M$ being a vector meson. We suspect that the chiral enhancement of scalar
meson production in factorization penguins is a property of factorization
of the local product rather than a general feature of all charming
penguin contributions.

\section{Fitting method}

We have attempted to fit the theoretical expressions for branching ratios
with the available data as averaged by the Heavy Flavour Averaging Group
\cite{HFAG}. Measured branching ratios for 18 channels are shown in table
\ref{BRs}. We take the measured branching ratios to be the mean of the
$B$ and $\overline{B}$ decays. For the two vector-vector channels $\phi K^*$
we multiply the measured branching ratios by the longitudinal polarization
as measured by BaBar~\cite{VVPol}.
CP asymmetries are not included in the fit. The
measurements are not always consistent between the experiments and the
errors are large. We therefore prefer to compare the measured results
with the predictions from the fit.
To economize in the number of soft QCD parameters we have not included
decay channels involving $\eta $ and $\eta^{\prime}$ mesons. These
amplitudes involve the mixing angle between the $(u\bar{u}+d\bar{d})$
and $s\bar{s}$ combinations. Also, in principle, there is mixing with
$c\bar{c}$ which, though small, could make a significant contribution
to decay modes through the enhanced quark decay modes
$b \rightarrow cq\bar{c}$.

For convenience we assign to each channel $(M_{1},M_{2})$ a number $\alpha $.
The statistical and systematic errors have been combined  \cite{HFAG}
into a single
error $\sigma_{\alpha}$. The systematic errors are in general small and
we ignore any correlations. We then form a $\chi ^{2}$ function
\begin{equation}
\label{chi2}
\chi^{2}(P_{i})  = \sum_{\alpha}[|Br_{\alpha}(P_{i})-
Br_{\alpha}(\text{exp})|
/\sigma_{\alpha}]^{2}  
 +\;\text{additional constraints}.
\end{equation}
$Br_{\alpha}(P_{i})$ are the theoretical branching ratios expressed in terms
of ten parameters $P_{i}$ which we take to be the three Wolfenstein CKM
parameters $\{A, \rho ,\eta \}$ and seven soft QCD parameters
$\{r^{\pi}_{\chi},r^{K}_{\chi},F_{\pi},F_{K},A_{\rho},A_{\omega},A_{K^{*}}\}$.
For the fit to the charming penguin model we introduce four more parameters:
the modulus and phase of $\overline{P_{1}}$ and
$\overline{P^{GIM}_{1}}$. In
this case we fix $A$ to the world average of 0.82 and keep
$r^{\pi,K}_{\chi}$ fixed at 1.0. The well established decay parameters
$\{f_{\pi},f_{K},f_{\rho},f_{\omega},f_{\phi},f_{K^{*}}\}$ are held at their
mean values and the Wolfenstein CKM parameter $\lambda $ is set to 0.2205.
The results are not very sensitive to the divergence parameters and we held
them fixed at $\rho_{H}=2e^{4.7i}$ and $X_{A}=1.85e^{2.86i}$, values suggested
by some preliminary investigations. Additional terms were
included in the $\chi^{2}$ to take into account experimental and theoretical
constraints from outside the data on $B$ decay branching ratios. We search
for a minimum of $\chi^{2}$ as a function of the $P_{i}$ using the
MINUIT \cite{minuit} program.

Next to the experimental error on the measurement we have to consider the
error from our assumptions on the QCD parameters that we do not fit for.
Table~\ref{SysVar} shows their central value and an estimate of
the allowed variation. First we performed the fit using experimental errors 
only. With the
best value we calculated a set of reference branching fractions. We then
varied the parameters according to Table~\ref{SysVar} while keeping the
value of the CKM parameters $\rho$ and $\eta$ fixed. For each parameter this
leads to a difference for each branching ratio. For every branching ratio
we sum the difference in quadrature and consider this to be the model
uncertainty. We add this uncertainty in quadrature to the experimental error
and repeat the fit.

\begingroup
\squeezetable
\begin{table*}
\caption{\label{BRs} Measured branching ratios Br(exp), experimental error
$\sigma $, best fit theoretical branching ratios Br(BBNS) and Br(CP)
for the Beneke \textit{et al.} model (BBNS) and charming penguin model (CP)
respectively, and the contribution to $\chi^{2}$ for various $B$ decay
channels. All branching ratios are in units of $10^{-6}$. For the $VV$
channels the predictions are for longitudinal polarization states only as
these decays are expected to be dominant.}
\begin{ruledtabular}
\begin{tabular}{cdddddd}
Decay & \multicolumn{1}{c}{Br(exp)} & \multicolumn{1}{c}{$\sigma $} &
\multicolumn{1}{c}{Br(BBNS)} & \multicolumn{1}{c}{$\chi^{2}$} &
\multicolumn{1}{c}{Br(CP)} & \multicolumn{1}{c}{$\chi^{2}$} \\
\hline
$\pi^{+}\pi^{-}$ & 4.8 & 0.5 & 5.0 & 0.2 & 4.9 & 0.0  \\
$\pi^{0}\pi^{-}$ & 5.6 & 0.9 & 4.0 & 2.0 & 5.9 & 0.0  \\
$\rho^{+}\pi^{-}$ & 25.4 & 4.2 & 24.4 & 0.1 & 23.3 & 0.2  \\
$\rho^{+}K^{-}$ & 16 & 5 & 7.3 & 2.7 & 11.1 & 0.7  \\
$\rho^{0}\pi^{-}$ & 9.4 & 2.0 & 9.5 & 0.0 & 9.6 & 0.0  \\
$\omega \pi^{-}$ & 6.4 & 1.3 & 7.4 & 0.5 & 6.4 & 0.0  \\
$\pi^{0}K^{-}$ & 12.9 & 1.2 & 13.0 & 0.0 & 12.9 & 0.0 \\
$\pi^{-}\overline{K}^{0}$ & 18.2 & 1.7 & 20.9 & 2.1 & 19.5 & 0.5  \\
$\pi^{-}\overline{K}^{*0}$ & 12.4 & 2.6 & 4.4 & 7.7 & 9.1 & 0.8  \\
$\omega K^{-}$ & 3.1 & 1.0 & 3.4 & 0.1 & 5.0 & 2.2  \\
$\phi K^{-}$ & 8.8 & 1.1 & 8.4 & 0.1 & 8.4 & 0.1  \\
$\phi K^{*-}$ & 5.4 & 2.4 & 8.9 & 1.0 & 9.0 & 1.6  \\
$\pi^{+}K^{-}$ & 18.5 & 1.0 & 18.7 & 0.1 & 19.0 & 0.2  \\
$\pi^{+}K^{*-}$ & 16 & 6 & 4.1 & 3.8 & 9.5 & 1.0  \\
$\pi^{0}\overline{K}^{0}$ & 10.4 & 1.4 & 7.4 & 3.9 & 7.1 & 4.6  \\
$\omega \overline{K}^{0}$ & 6.5 & 1.7 & 2.4 & 5.2 & 4.4 & 1.2  \\
$\phi \overline{K}^{0}$ & 8.4 & 1.6 & 7.8 & 0.2 & 7.8 & 0.2  \\
$\phi \overline{K}^{*0}$ & 7.2 & 1.7 & 8.3 & 0.3 & 8.3 & 0.2  \\
\end{tabular}
\end{ruledtabular}
\end{table*}
\endgroup

\begin{table}
\caption{\label{SysVar} Value of the parameters used with their variation.}
\begin{ruledtabular}
\begin{tabular}{cdd}
Parameter & \multicolumn{1}{c}{Central Value} & \multicolumn{1}{c}{Variation} \\ 
\hline
$f_{\pi}$ & 0.1307 & \pm 0.00046 \\
$f_{K}$ & 0.1598 & \pm 0.00184 \\
$f_{\rho}$ & 0.216 & \pm 0.005\\
$f_{\omega}$ & 0.194 & \pm 0.004\\
$f_{K^*}$ & 0.216 & \pm 0.010\\
$f_{\phi}$ & 0.233 & \pm 0.004\\ 
$\lambda$ & 0.2205 & \pm 0.0010 \\
$\tau_{B^0} / \tau_{B^{\pm}} $ & 1.081 & \pm 0.015 \\
$\mu$ & 0.5 m_{b} & m_b \\
$|\rho_H|$ & 2.0 & \pm 1.0 \\
arg$(\rho_H)$ & 4.7 & 1.6 \\
$|X_A|$ & 1.85 & \pm 1.0 \\
arg$(X_A)$ & 2.86 & 3.7 \\
$r_{\chi}^{\pi,K}$ & 1.0 & \pm 0.2 \text{(CP only)} \\
$A_{\text{Wolf}}$ & 0.82 & \pm 0.05 \text{(CP only)} \\
$\alpha^M_n$ & 0 & \multicolumn {1}{c}{Table IV (L\"{u} and Yang)}\\
\end{tabular}
\end{ruledtabular}
\end{table}

The theoretical branching ratios and the contributions of the individual
channels to $\chi^{2}$ based upon these best fit values are given in table
\ref{BRs}. The best fit parameter values are shown in table \ref{fit}
together with our estimates of the errors. These errors are of course
highly correlated. Plots of the error matrix ellipse for the Wolfenstein
parameters $\bar{\rho} $ and $\bar{\eta} $ are shown in
figures \ref{fig:UT1} and \ref{fig:UT2}. The CKM angles are
$\alpha = (78 \pm 9)^{\circ},\;\beta = (22 \pm 2)^{\circ},
\;\gamma =(80 \pm 7)^{\circ}$ for the BBNS model and
$\alpha = (63 \pm 7)^{\circ},\;\beta = (21 \pm 2)^{\circ},
\;\gamma =(96 \pm 6)^{\circ}$ for the CP model. These can be compared with
the world averages of \cite{CKMfitter}
$\alpha = (96 \pm 13)^{\circ},\;\beta = (23.3 \pm 1.5)^{\circ},
\;\gamma =(61 \pm 12)^{\circ}$. All errors are one standard deviation.
For both models the
results in table \ref{fit} for the values of the various form factors lie
within the spread of theoretical estimates. The $\chi^2$/dof is 34.5/17 for
the BBNS fit and 14.5/13 for the charming penguins.

\begingroup
\squeezetable
\begin{table*}
\caption{\label{fit} Best fit values and one standard deviation errors
for the Beneke \textit{et al.} (BBNS) and charming penguin (CP) models.
$F_{\pi,K}$ and $A_{\rho,\omega ,K^{*}}$ are the transition form factors
$F^{B \rightarrow h}(0)$ for $P$ and $V$ mesons respectively,
and $r^{\pi,K}_{\chi}$ are the chiral
enhancement factors which are nominally power suppressed but are $O(1)$
in practice.}
\begin{ruledtabular}
\begin{tabular}{cddddd}
Model & \multicolumn{1}{c}{$F_{\pi}$} & \multicolumn{1}{c}{$F_{K}$} &
\multicolumn{1}{c}{$A_{\rho}$} & \multicolumn{1}{c}{$A_{\omega}$} &
\multicolumn{1}{c}{$A_{K^{*}}$}  \\
\hline
BBNS & 0.244\pm 0.038 & 0.369\pm 0.031 & 0.344\pm 0.098  &
0.300\pm 0.094  & 0.321\pm 0.136  \\
CP  & 0.291\pm 0.022 & 0.349\pm 0.077 & 0.320\pm 0.065  &
0.298\pm 0.072  & 0.309\pm 0.117  \\
&&&&&  \\
& \multicolumn{1}{c}{$r^{\pi}_{\chi}$} & \multicolumn{1}{c}{$r^{K}_{\chi}$} &
\multicolumn{1}{c}{$A$} & \multicolumn{1}{c}{$\bar{\rho}$}  &
\multicolumn{1}{c}{$\bar{\eta}$ }  \\
\hline
BBNS & 1.09\pm 0.24 & 1.24\pm 0.17 & 0.813\pm 0.045  &
0.068\pm 0.071  & 0.383\pm 0.090  \\
CP  & 1.0  & 1.0  & 0.82  & -0.044\pm 0.112  & 0.397\pm 0.050  \\
&&&&&  \\
& \multicolumn{1}{c}{$|D|$} & \multicolumn{1}{c}{Arg($D$)} &
\multicolumn{1}{c}{$D^{GIM}$} & \multicolumn{1}{c}{Arg($D^{GIM}$)} \\
\hline
CP & 0.068\pm 0.007  & 1.32\pm 0.10  & 0.32\pm 0.14
& 1.00\pm 0.27  & \\
\end{tabular}
\end{ruledtabular}
\end{table*}
\endgroup

\section{Discussion and Conclusions}

\begin{figure}[htb]
\includegraphics[width=0.5\linewidth]{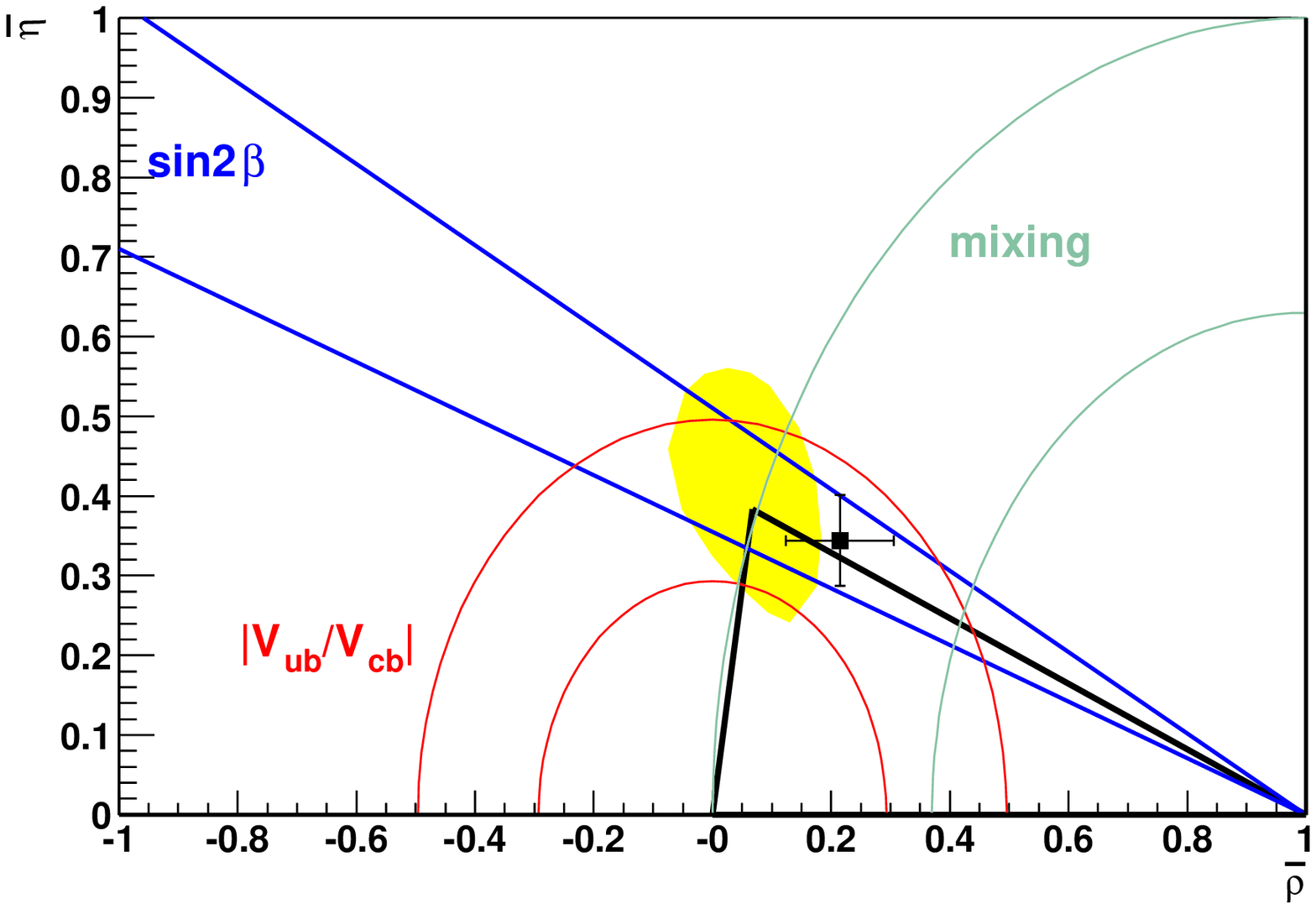}
\caption{\label{fig:UT1}
Result for the unitarity triangle fit in the BBNS model.
The shaded area shows the $3\sigma$ allowed region
for the apex of the unitarity triangle. The data point shows the fit result
from the unitarity triangle fit from other measurements taken 
from \textit{CKMfitter}\cite{CKMfitter}.}
\end{figure}

\begin{figure}[htb]
\includegraphics[width=0.5\linewidth]{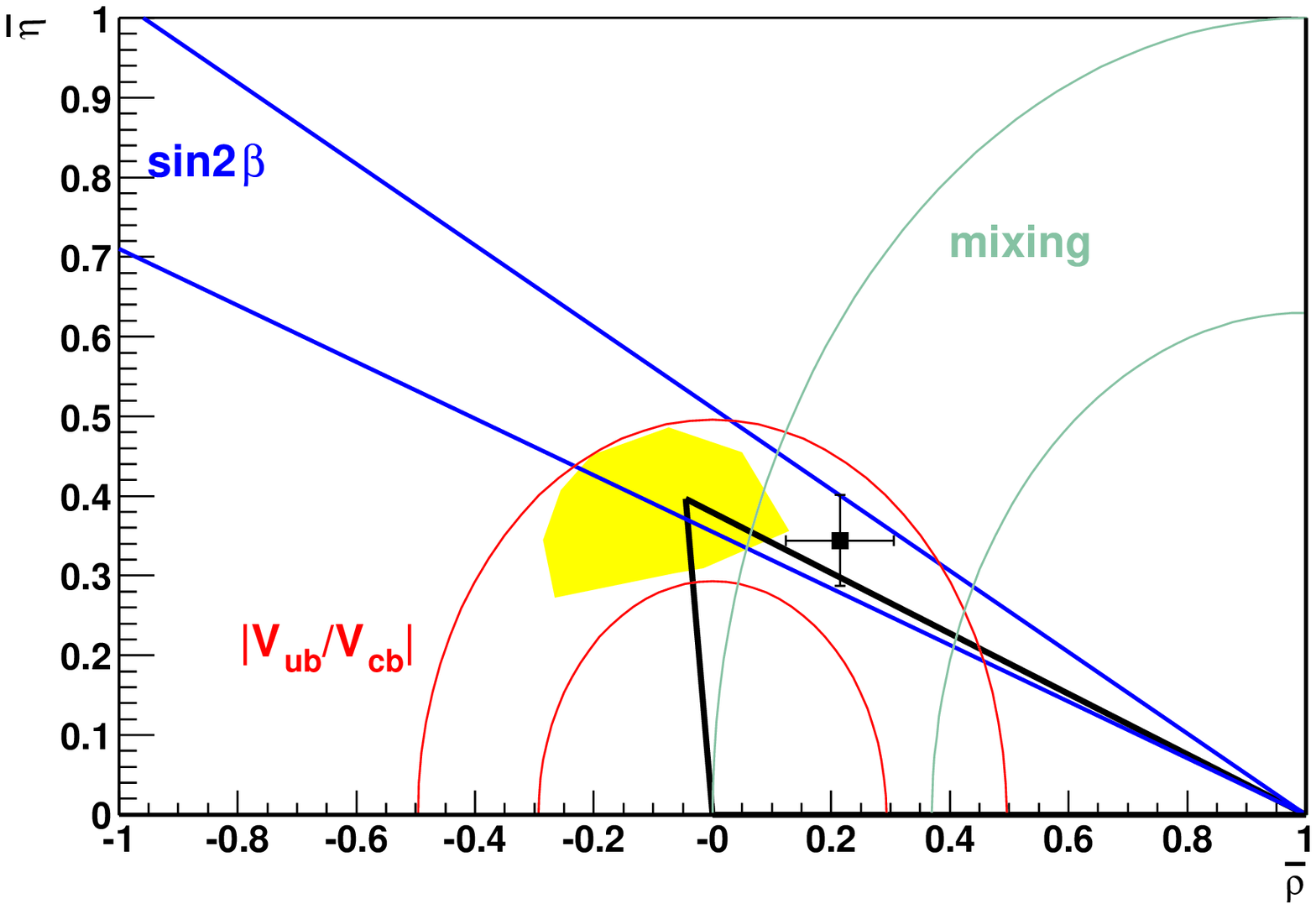}
\caption{\label{fig:UT2}
Result for the unitarity triangle fit in the charming penguin model.
The shaded area shows the $3\sigma$ allowed region
for the apex of the unitarity triangle. The data point shows the fit result
from the unitarity triangle fit from other measurements taken 
from \textit{CKMfitter}\cite{CKMfitter}.}
\end{figure}

The first conclusion is that the factorization approach works quite well.
Most of the branching ratios in table \ref{BRs} are predicted correctly
by both models. The fitted parameters in table \ref{fit} look reasonable
for both fits. The $\chi^{2}$ for the charming penguin fit is
significantly better, due mainly to the poor fit of the BBNS model for
the decay modes involving the $K^{*}$ meson. Also, the experimental value
for $\omega K^{0}$ is not easily accomodated within the BBNS model. Both
models have a problem in fitting the $\pi^{0}K^{0}$ mode.
Figures \ref{fig:UT1} and \ref{fig:UT2}
show the position of the apex of the unitarity triangle for both fits.
The results are consistent with each other for the angle $\gamma $ but
give a larger value than that suggested by the CKM Fitting Group. The
angle $\beta $ agrees well for both fits.

Our theoretical best-fit values for those $CP$ asymmetries that have been
measured are shown in table \ref{asym}.
The direct $CP$-violating parameter $A_{\text{CP}}$ for the decay
channel $M_{1}M_{2}$ is
\begin{equation}
\label{acp}
A_{\text{CP}}=\frac{\Gamma (\bar{B} \rightarrow \bar{M}_{1}\bar{M}_{2})
-\Gamma (B \rightarrow M_{1}M_{2})}
{\Gamma (\bar{B} \rightarrow \bar{M}_{1}\bar{M}_{2})
+\Gamma (B \rightarrow M_{1}M_{2})}
\end{equation}
where $\bar{B}=b\bar{u}$ or $b\bar{d}$. The definitions of the other
$CP$-violating parameters can be found in \cite{babarcp}.
Regarding these asymmetries
it is too early to reach a conclusion. In
many cases the different experiments disagree, in others the errors are so
large that a meaningful discrimination is not possible. The
theoretical asymmetries are
very sensitive to the parameters and to the different models and, with
improved statistics, could become the final test of factorization.
In table \ref{extra} we show our predictions for branching ratios and
$CP$ asymmetry $A_{\text{CP}}$ for some channels not included in the
fit \footnote{Experimental results for two of these decay modes are now
available from the BaBar and Belle experiments. They find \cite{HFAG}
$B^{0} \rightarrow \pi^{0} \pi^{0} = (1.9 \pm 0.5) \times 10^{-6}$ and
$B^{-} \rightarrow \rho^{-} \pi^{0} = (11.0 \pm 2.7) \times 10^{-6}$. The
former is on the high side of our predictions, particularly for the
BBNS model, the latter is in good agreement with the prediction from the
CP model but less so with that of the BBNS model}.
We include some $VV$ channels that are currently under investigation.
For these channels we have assumed that the decays are to longitudinally
polarized vector mesons as it is expected \cite{Cott,Yang} that decays to
the other polarization states will be suppressed by at least a factor of
$(m_{V}/m_{B})^{2}$.

In their most recent work Beneke and Neubert \cite{BenekePV} give formulae
for the weak annihilation functions $A^{i,f}_{1,3}(P,V)$ that, in addition
to the terms given in (\ref{A-cofs}), include terms containing a
parameter $r^{V}_{\chi}$. This parameter $r^{V}_{\chi}$ has a similar
origin to $r^{P}_{\chi}$ and, like $r^{P}_{\chi}$, is suppressed by a
power of $\Lambda_{\text{QCD}}/m_{b}$ but, unlike $r^{P}_{\chi}$, is not
chirally enhanced. We only became aware of \cite{BenekePV} after completion
of this study. To check for the effects of these terms we have modified
our program to include the $r^{V}_{\chi}$ contribution to annihilation and
have reminimized $\chi ^{2}$. For the charming penguin model we found the
best fit occurred with $X_{A}=1.09 \exp (2.79i)$ and that there were very
small changes in the best fit parameters of table \ref{fit} and the results
of tables \ref{BRs}, \ref{asym} and \ref{extra}. For the BBNS model we
found similar results but the overall fit was better in that the $\chi^{2}$
was reduced from 34.5 to 27.5.

Finally, the results presented here are slightly different from preliminary
numbers presented earlier~\cite{moriond03}, for three reasons.
The Heavy Flavour Averaging Group has updated some of the results and
included a new branching fraction ($\rho^{\pm}K^{\mp}$).
We have also corrected
the vector-vector channels $K^*\phi$ for the effect of polarization. Finally
we now include the systematic uncertainties in the fit.

\begin{table}
\caption{\label{asym} Measurements and theoretical best fit values for
$CP$ asymmetries.
Only statistical errors are shown.}
\begin{ruledtabular}
\begin{tabular}{crrrr}
$A_{\text{CP}}$ & \multicolumn{1}{c}{BaBar} & \multicolumn{1}{c}{Belle} &
\multicolumn{1}{c}{BBNS} & \multicolumn{1}{c}{CP} \\
\hline
$K^{+}\pi^{-}$ & $-0.102\pm 0.05$  & $-0.07\pm 0.06$  & 0.00  & -0.08  \\
$K^{+}\pi^{0}$ & $-0.09\pm 0.09$  & $0.23\pm 0.11$  & 0.06  & -0.02  \\
$K^{0}\pi^{+}$ & $-0.17\pm 0.10$  & $0.07\pm 0.09$  & 0.01  & 0.11  \\
$\pi^{+}\pi^{0}$ & $-0.03\pm 0.18$  & $-0.14\pm 0.24$  & 0.00  &  0.00\\
$\rho^{+}\pi^{-}$ & $-0.22\pm 0.08$  & & $-0.03$   & $-0.03$  \\
$\rho^{+}K^{-}$ & $0.28\pm 0.17$ & & $ 0.10$  & $-0.40$  \\
&&&&  \\
$C_{\pi\pi}$ & $-0.30\pm 0.25$  & $-0.77\pm 0.27$  & 0.01  & $-0.23$  \\
$S_{\pi\pi}$ & $0.02\pm 0.34$  & $-1.23\pm 0.41$  & -0.20  & 0.25  \\
$C_{\rho\pi}$ & $0.36\pm 0.15$  & & 0.03  & $-0.31$  \\
$S_{\rho\pi}$ & $0.19\pm 0.24$  & & 0.38  & 0.65  \\
$\Delta C_{\rho\pi}$ & $0.28\pm 0.19$  & & 0.08  & 0.31  \\
$\Delta S_{\rho\pi}$ & $0.15\pm 0.25$ & & 0.02   & 0.11  \\
$C_{\phi K^{0}_{s}}$ & $-0.80\pm 0.38$ & $0.56\pm 0.41$  & $-0.01$
& $-0.19$  \\
$S_{\phi K^{0}_{s}}$ & $-0.18\pm 0.51$  & $-0.73\pm 0.64$  & 0.73  & 0.62  \\
\end{tabular}
\end{ruledtabular}
\end{table}

\begin{table}
\caption{\label{extra} Predicted branching ratios (in units of $10^{-6}$) and
$CP$ asymmetries for some channels not included in the fit.
For the $VV$
channels the predictions are for longitudinal polarization states only as
these decays are expected to be dominant.
The predictions are for the central
values of the model fits with $ 1 \sigma $ errors estimated by sampling
the parameter error matrix.}
\begin{ruledtabular}
\begin{tabular}{crrrr}
& \multicolumn{2}{c}{Mean Branching Ratio}
& \multicolumn{2}{c}{$A_{\text{CP}}$}   \\
\hline
Decay & \multicolumn{1}{c}{BBNS}  &  \multicolumn{1}{c}{CP}  &
\multicolumn{1}{c}{BBNS}  &  \multicolumn{1}{c}{CP} \\
\hline
$\pi^{0}\pi^{0}$ & $0.5\pm 0.1$  & $1.1\pm 0.2$  & $0.59\pm 0.08 $
& $-0.63\pm 0.15$ \\
$\rho^{0}\rho^{0}$ & $0.5\pm 0.1$  &  $1.2 \pm 0.3$ & $0.69 \pm 0.16$ &
$-0.66 \pm 0.07$  \\
$\pi^{0}\rho^{-}$ & $8.4\pm 0.7$ & $11.9 \pm 1.4$ & $0.04 \pm 0.01$  &
$0.06 \pm 0.01$  \\
$\rho^{0}\rho^{-}$ & $19.0 \pm 3.1$ & $17.8 \pm 3.3$ & $0.00 \pm 0.00$ &
$0.00 \pm 0.00$  \\
$\rho^{0}K^{+}$ &  $2.0 \pm 0.6$ & $4.6 \pm 0.5$ & $ 0.00 \pm 0.09$ &
$-0.01 \pm 0.64$  \\
$\rho^{0}K^{*+}$ & $4.6 \pm 1.3 $ & $ 5.2 \pm 0.7 $ & $ 0.33 \pm 0.06$ &
$ -0.35\pm 0.06$  \\
\end{tabular}
\end{ruledtabular}
\end{table}

\bibliographystyle{prsty}

\end{document}